\documentclass[12pt,preprint]{aastex}

\shorttitle{Extinction in Star-Forming Regions}
\shortauthors{McClure}
\usepackage{rotating}
\usepackage{lscape}

\begin{document}

\title{Observational $5-20\mu$m Interstellar Extinction Curves Toward Star-Forming Regions Derived from Spitzer IRS Spectra}


\author{M. McClure\altaffilmark{1,2}}

\altaffiltext{1}{Department of Astronomy, University of Michigan, Ann Arbor, MI 48104; melisma@umich.edu}
\altaffiltext{2}{Department of Physics and Astronomy, University of Rochester, Rochester, 
NY 14627}

\begin{abstract}
Using \emph{Spitzer} Infrared Spectrograph observations of G0--M4 III stars behind dark clouds, I construct $5-20\mu$m empirical extinction curves for $0.3\leq A_K<7$, which is equivalent to $A_V$ between $\approx$ 3 and 50.  For $A_K<1$ the curve appears similar to the \citet{mathis90} diffuse interstellar medium extinction curve, but with a greater degree of extinction.  For $A_K>1$, the cuve exhibits lower contrast between the silicate and absorption continuum, developes ice absorption, and lies closer to the \citet{wd01} $R_V=5.5$ case B curve, a result which is consistent with that of \citet{flaherty07} and \citet{chiar07}. Recently work using \emph{Spitzer} Infrared Array Camera data by \citet{chapman08} independently reaches a similar conclusion, that the shape of the extinction curve changes as a function of increasing $A_K$.  By calculating the optical depths of the $9.7\mu$m silicate and 6.0, 6.8, and 15.2 $\mu$m ice features, I determine that a process involving ice is responsible for the changing shape of the extinction curve and speculate that this process is coagulation of ice-mantled grains rather than ice-mantled grains alone.
\end{abstract}


\section{Introduction}
Extinction along the line of sight to a young stellar object must be accounted for when considering the nature of that object.  Discrepancies between the diffuse interstellar medium (DISM) extinction curve and observations of dark clouds, where young stellar objects form, have been noted at UV and visible wavelengths \citep[][and references therein]{sm79}.  These variations were successfully modeled by parametrizing the extinction curve with $R_V$, the total-to-selective extinction \citep[][hereafter CCM89 and WD01]{ccm89,wd01}, which varies from an average value of 3.1 in the DISM to about 5 in dense clouds \citep{whittet01}.  Most of the resulting curves are remarkably consistent at wavelengths longer than $0.9\mu$m regardless of the assumed $R_V$ (see CCM89 and WD01 Case A), the exception being a WD01 $R_V=5.5$ curve in which the maximum grain size is $10\mu$m.  This `Case B' extinction curve is considerably higher than either the other WD01 or CCM89 extinction curves at wavelengths longer than $3.0\mu$m.  More recent work has demonstrated that extinction towards dense molecular clouds \emph{is} different from DISM extinction over the $3-24\mu$m micron range, particularly over the $3-8\mu$m region \citep{indebetouw05, flaherty07, chapman08} and the shape of the $9.7\mu$m silicate feature \citep[][hereafter C+07]{chiar07}.  Many of the nearby star-forming regions (e.g. Ophiuchus, Orion) have a large fraction of highly extinguished members.  For $A_V<12$ (equivalent to $A_K=1.5$) past the breaking point of the local ISM correlation in dark clouds (C+07), the sample size for these regions is drastically reduced.  

To analyze a full sample from these regions, a new extinction curve is required.  Using \emph{Spitzer} Infrared Spectrograph (IRS) \citep{houck04} observations of stars with known intrinsic spectra behind various dense molecular clouds, I determine the the spectrum of extinction from $5-20\mu$m for 31 lines-of-sight.  While any curves derived for a particular line of sight are unlikely to be universal, the median curves for specific ranges of extinction should characterize the shape of the extinction curve for dark clouds.

\section{Sample and Analysis}

I began my analysis with 28 of the G0--M4 III stars discussed in C+07 that lie behind the Taurus, Chameleon I, Serpens, Barnard 59, Barnard 68, and IC 5146 molecular clouds.  Most of these have been observed with some combination of the \emph{Spitzer} IRS low-resolution ($\lambda$/$\Delta\lambda$  = 60-120) short-wavelength (SL; $5.2-14\mu$m) and long wavelength 2nd order (LL2; $14.0-21.3\mu$m) and the high-resolution ($\lambda$/$\Delta\lambda$ = 600) short-wavelength (SH; $10-19\mu$m) modules.  With the exception of HD 29647, a B8III star which has yet to be observed with the IRS, these background stars have spectral types in the range G0--M4 and should have little intrinsic emission in the mid-infrared, as long as they are not supergiants, which are quite rare.  One of these background stars, CK 2, was serendipitously observed with the 1$^{st}$ order of the long-wavelength, low-resolution module (LL1; $20-35\mu$m) as part of a GTO program in Serpens.  I supplemented this list of objects with three other background K and M III stars from \citet{shenoy08} that have been observed with the IRS.  The final list follows, with the AOR numbers, in Table 1.  With the exception of background emission subtraction in SL and LL, which was done from the opposite nod position in most cases, I extracted these spectra using the method described in \citet{furlan06}.  Two of the objects, Elias 3 and Elias 13 were observed with SL and SH, but SH was clearly contaminated by sky, as indicated by the excess flux above the level of SL and a different slope over the $10-14\mu$m range, so we only used SL for these observations. I also declined to use Elias 9, as it was only observed with SH.  

The line-of-sight extinction in a particular band can be determined from a color excess.  Taking \emph{H} and \emph{K$_S$} (henceforth \emph{K}) photometry from 2MASS \citep{cutri03} and using the most recent online update of the \citet{carpenter01} color transformations to convert the intrinsic luminosity class III colors from \citet{bb88} into the 2MASS system, I calculated the color excess $E(H-K)$, assuming a spectral type of K2 for objects with spectral types listed as G0--M4.   Uncertainties in the $E(H-K)_{intrinsic}$ include the independent contributions of uncertainties in 2MASS colors and transformed intrinsic colors.  For objects with spectral types in the range G0--M4, the uncertainy in $(H-K)_{intrinsic}$ was taken to be the difference between $(H-K)_{intrinsic}$ of an M4 giant and of a K2 giant, since this was greater than the difference between a K2 and G0 giant, resulting in $(H-K)_{intrinsic}$ of $0.146\pm0.134$.  The choice of K2 to represent the spectral types of objects with a range of G0--M4 may affect the precise value of extinction, but it will have less effect on the shape of the extinction curve over the IRS spectrum, since most photospheres follow a Rayleigh-Jeans tail at $\lambda>5\mu$m.  For objects with a known spectral type, the uncertainty was calculated by assuming a range of $\pm$ one subgroup.  The color excess is defined as $E(H-K)=A_H-A_{K}$, from which the extinction, $A_K$, for these stars was calculated using the following expression:

\begin{equation}
A_{K}=\frac{\left[E(H-K\right)]}{\frac{A_H}{A_K}-1}
\end{equation}

\noindent and uncertainties in $A_K$ determined via error propagation. I took $A_H/A_K=1.56$ from the \citet{mathis90} extinction law, interpolated to the 2MASS H and K wavelengths.  The slopes of extinction laws in the literature are relatively uniform over \emph{JHK} (CCM89; WD01).

Two of the stars required special consideration.  SSTc2dJ182852.7+02824 was undetected in the \emph{K} band of 2MASS, but a magnitude at \emph{K} was given by \citet{denis05}.  I calculated $A_K$ as above but without converting the DENIS \emph{K} into the 2MASS system and note that there is a small additional uncertainty associated with this extinction (which is much smaller than the uncertainty of the spectral type).  B59-bg1 was undetected at \emph{H}, so this extinction is a lower limit to the true value. 

Next I interpolated a model photosphere \citep{castelli97} of appropriate spectral type to the resolution of the 2MASS data and the IRS spectrum for each object, then extinction corrected the 2MASS \emph{K} flux using $A_K$.  $A_{\lambda}$ can then be found using the following expression, where $F_{K_{corrected}}$ is used to scale the photosphere appropriately:

\begin{equation}
A_{\lambda}=2.5log(\frac{F_{K_{corrected}}}{F_{K_{photosphere}}}\frac{F_{\lambda_{photosphere}}}{F_{\lambda_{spectrum}}})
\end{equation}

\noindent I also calculated the peak optical depth of the $9.7$ (silicates), 6.0 (H$_2$O ice), 6.8 (`methanol' ice), and $15.2\mu$m (CO$_2$ ice) absortption features for each source with, e.g, $\tau_{9.7}$=-ln($\frac{F_{9.7_{spectrum}}}{F_{9.7_{continuum}}}$), where the continuum was the photosphere scaled to $F_{K_{corrected}}$.

\section{Results and Discussion}

\subsection{New extinction curves}
\begin{figure}
\epsscale{0.7}
\plotone{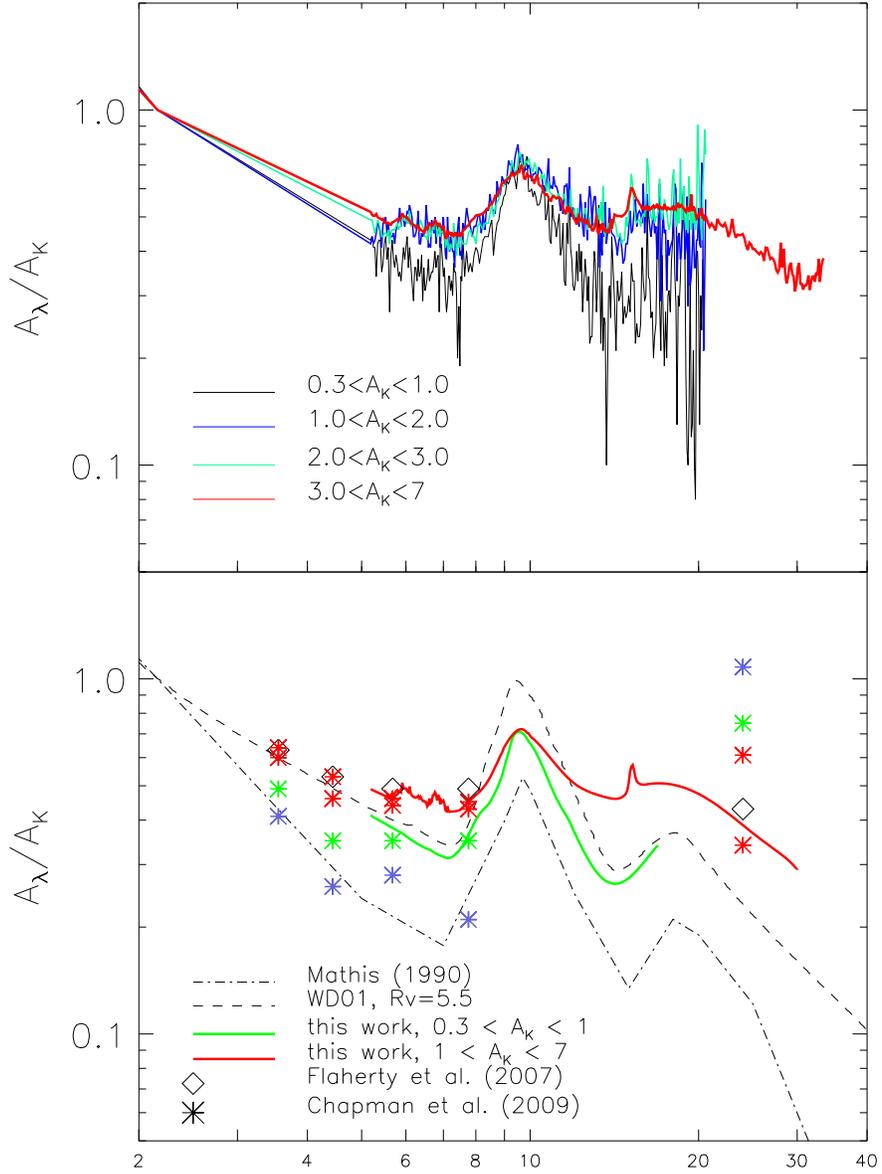}
\caption{\emph{Top}: Medians of extinction curves toward stars with $A_K$ of 0.28-1.0 (black), 1.0-2.0 (dark blue), 2.0-3.0 (cyan), and 3.0-7 (red) mags.  \emph{Bottom}: New extinction curves for $0.3\leq A_K<1$ (green, solid) and $1\leq A_K<7$ (red, solid) (see text for details). The \citet{mathis90} and WD01 $R_V$=5.5 case B extinction curves are also plotted (black, dash-dotted and black, dashed, respectively).  The \citet{flaherty07} IRAC and MIPS extinction law is plotted with open diamonds. For comparison, the new \citet{chapman08} extinction curves, which cover the same wavelength range, are plotted for $0<A_K\leq 0.5$ (blue asterisks), $0.5<A_K\leq 1$ (green asterisks), $1<A_K\leq 2$ (red asterisks, lower), and $2\leq A_K$ (red asterisks, higher).\label{extcompare}}
\end{figure}

To analyze these extinction curves, I first normalized $A_{\lambda}$ to $A_K$ for each object, then separated the objects into groups according to their level of extinction.  Eleven objects were in the $0.3\leq A_K<1.0$ category, ten in $1.0\leq A_K<2.0$, six in $2.0\leq A_K<3.0$, and four in $3.0\leq A_K<7$.  Of these objects, all had $5-14\mu$m spectra and the majority had coverage from $14-20\mu$m, either in SL or SH,  with one of those having additional coverage in LL1.  The medians of all four groups are shown in Fig. \ref{extcompare} (top); the means were very similar to the medians, but with poorer signal-to-noise.  The groups with $1.0<A_K<7$ have extremely similar medians, with higher extinction from $6-8\mu$m and $10-20\mu$m, a $9.7\mu$m silicate feature with a wider long wavelength wing, and ice features that become more pronounced as $A_K$ increases, while the median of the $A_K<1.0$ bin is lower than the rest with more pronounced silicate features and no ice features.  I took the mean of the medians of the three most extinguished groups, and kept the median of the least extinguished group as it was.  To create a smooth extinction curve, I fit 2$^{nd}-4^{th}$ degree polynomials to the continuum regions and silicate features of both extinction curves, combined the polynomials, and inserted the mean ice features into the polynomial fit for the highest $A_K$ groups, resulting in two smooth extinction curves, one for $0.3\leq A_K<1.0$ and one for $1.0\leq A_K$. 

To create composite extinction curves with my data and those from the literature, I renormalized the \citet{mathis90} and WD01 $R_V$ = 5.5 Case B extinction curves to $A_K$.  Plotting the polynomial fitted extinction curves against both the \citet{mathis90} and WD01 curves, I noticed that the WD01 curve parallels the $0.3\leq A_K<1.0$ curve from $16\mu$m and longward, and it roughly matches the slope of the $1.0\leq A_K$ extinction curve beyond $30\mu$m.  Consequently, I scaled the \citet{wd01} curve up and appended it to the polynomial fit curves past approximately $16\mu$m and $30\mu$m.  Since my exinction curves were already normalized to \emph{K} band of the \citet{mathis90} curve, I prefixed my new curves with the the 3.6 and $4.5\mu$m extinctions from \citet{flaherty07} and with the Mathis curve up to $2.3\mu$m, assuming $R_V$ of 5.0 for $\lambda<$0.9$\mu$m.  The portions of the curves derived soley in this work are compared with other curves from the literature in Figure \ref{extcompare} (bottom) and the final, composite curves are tabulated in Table \ref{exttab}.  

Comparing the curves in the bottom panel of Figure \ref{extcompare}, it appears that there is real variation in the shape of the extinction curve as a function of $A_K$ in the mid-infrared.  The $0.3\leq A_K<1.0$ extinction has a similar overall slope to the \citet{mathis90} curve but has higher extinction all around. Originally, I had subdivided the objects with $0.3\leq A_K<1.0$ into two groups, and the lower group appeared even closer to the \citet{mathis90} curve, but the uncertainties on both $A_K$ and the poor signal to noise in some of the spectra in that subset necessitated using a larger range of objects (and hence $A_K$) to construct a good polynomial fit to the curve.  The $1.0\leq A_K$ curve is much higher than the \citet{mathis90}, WD01, or the $0.3\leq A_K<1.0$ curve but is consistent with the results of \citet{indebetouw05} and \citet{flaherty07}, which were based on \emph{Spitzer} Infrared Array Camera (IRAC) and Multiband Imaging Photometer (MIPS) photometry.  Significantly, the extinction over the $9.7-20\mu$m region is also higher than \emph{both} the \citet{mathis90} curve and the $0.3\leq A_K<1.0$ curve; in fact, it is almost flat.  This result, that the extinction curve transitions from a shape similar to the DISM \citet{mathis90} curve to a higher, flatter extinction, was independently derived from IRAC photometry for several molecular clouds by \citet{chapman08}, whose extinction curves roughly match mine over the $5-8\mu$m region for $A_K>0.5$.  The range of $24\mu$m extinctions given by both \citet{flaherty07} and \citet{chapman08} are consistent with a the flat slope of the $1\leq A_K$ extinction curve derived here from $8-24\mu$m.  Unfortunately, none of the data with $A_K<1$ extended beyond $20\mu$m for comparison.  

\subsection{Shape of the $9.7\mu$m silicate feature}
\begin{figure}
\plotone{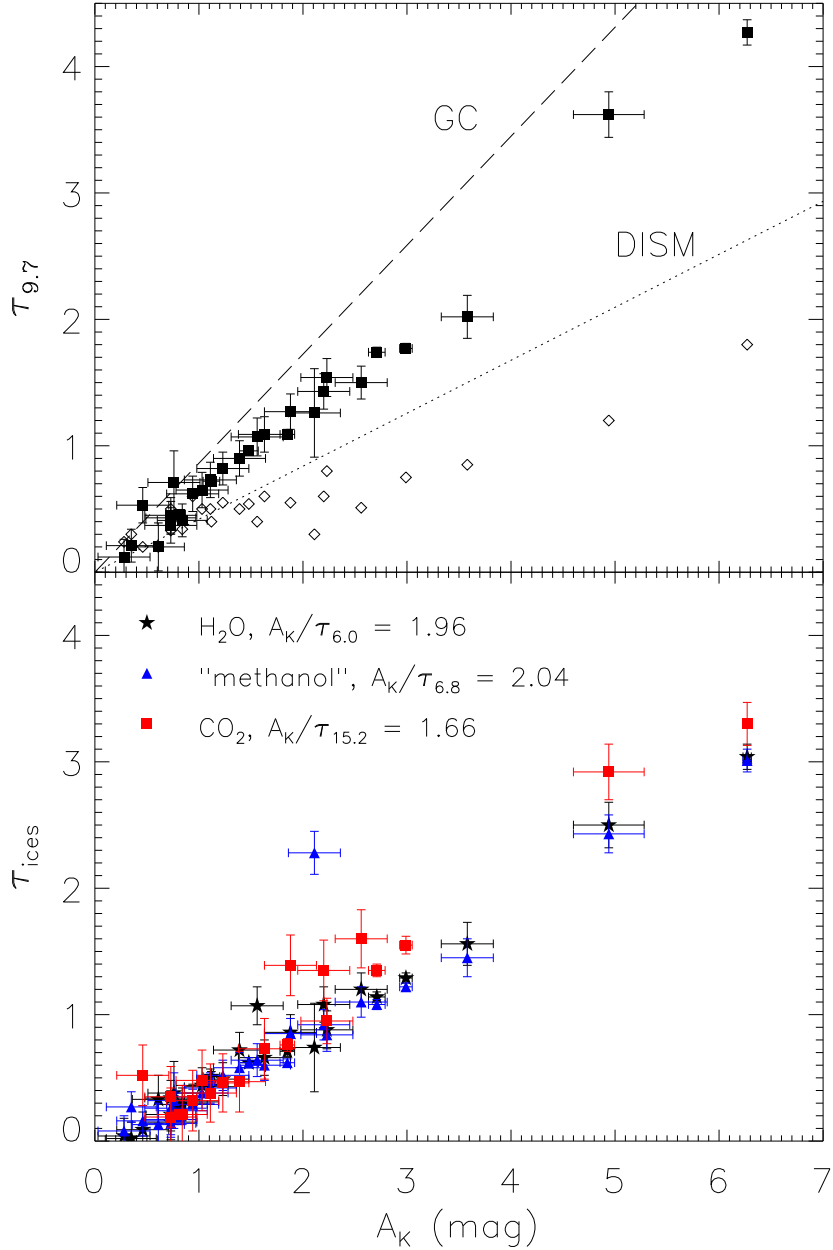}
\caption{$A_K$ vs $\tau$.  \emph{Top}: $\tau_{9.7}$ from this work (filled squares with error bars).  Stars with $\tau_{9.7}$ from C+07 (open diamonds) are plotted using the $A_K$ we derive here.  Dotted and dashed lines represent the linear relationships between $A_K$ and $\tau_{9.7}$ for the diffuse ISM (DISM; $A_V=18.5$, i.e. $A_K=2.39$) and galactic center (GC, $A_V=9$, i.e. $A_K=1.16$), respectively (WD01, references therein). \emph{Bottom}: $\tau_{ices}$, for H$_2$O, `methanol', and CO$_2$.  \label{Avtaus}}
\end{figure}

In addition to changes in the slope of the extinction curve, the silicate features change shape with increasing $A_K$ as well.  The amplitude relative to the 7 and $14\mu$m regions of the $9.7\mu$m silicate feature in the $0.3\leq A_K<1.0$ curve is somewhat smaller than in the \citet{mathis90} curve or the WD01 Case B curve.  As $A_K$ passes 1 magnitude, the amplitude decreases even further and the longer wavelength wing begins to broaden.  Noticeably, the $18\mu$m silicate feature is considerably wider and flatter in the $1.0\leq A_K$ curve than in the literature curves. That the amplitude of the $9.7\mu$m silicate feature relative to the rest of the curve changes as a function of $A_K$ is similar to the findings of C+07.  Plotting $A_K$ against $\tau_{9.7}$, the optical depth increases linearly as a function of the extinction (Fig. \ref{Avtaus}).  Taking a least squares fit to the data, I find that the linear relationship is $A_K/\tau_{9.7}$ = 1.48 $\pm$ 0.02 with R = 0.989, which is equivalent to $A_V/\tau_{9.7}$ = 11.46 if $R_V=5.0$ (see footnote to Table \ref{SpT_all} for conversion factor).  I do not find the same break in the relationship between $\tau_{9.7}$ and $A_K$ at $A_K=1.5$, equivalent to $A_V=10-12$, that C+07 do.  Additionally, their data (open diamonds, Fig. \ref{Avtaus}), are much lower than mine.  This discrepancy is caused by the difference in how we calculate our continuua.  I take the continuua to be the photospheres scaled to the extinction corrected \emph{K}-band fluxes, while C+07 take theirs from 2$^{nd}$ or 3$^{rd}$ degree polynomial fits to regions of silicate-free continuum emission at $5.2-7\mu$m and $13.5-15\mu$m.  As a result, I am measuring the \emph{total} optical depth at $9.7\mu$m, while the C+07 data denote the $9.7\mu$m optical depth \emph{in excess} of the adjacent continuum optical depth.  

\begin{figure}
\epsscale{1.1}
\plotone{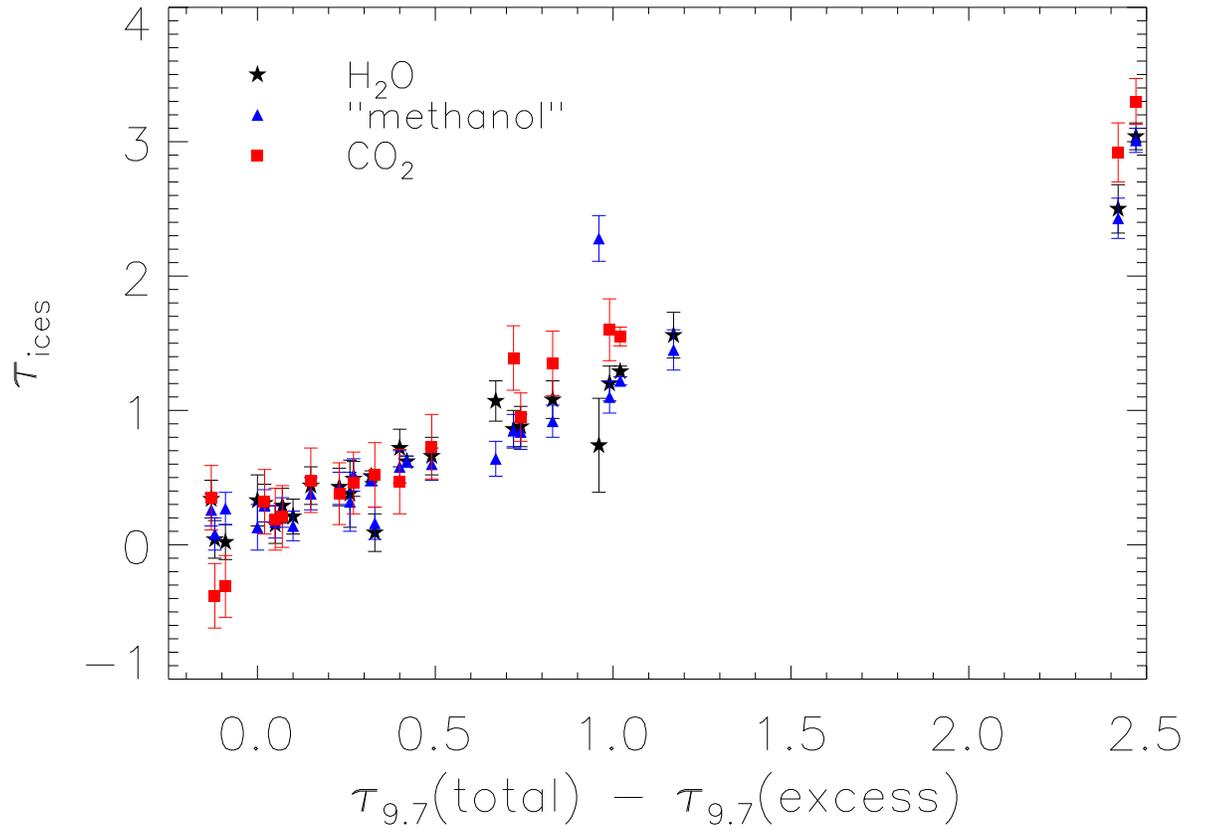}
\caption{Correlation between the optical depth of the $9.7\mu$m continuum extinction (the difference between $\tau_{9.7}$ measured here and the C+07 $\tau_{9.7}$ values) and the optical depths of the three ice features.  All three ice species correlate with $R>0.97$. \label{excess}}
\end{figure}

At higher $A_K$, if the effects of grain growth contribute to the extinction, one expects the silicate profile to broaden to longer wavelengths and extinction from scattering to be important, which could affect the $13.5-15\mu$m region used to anchor the polynomial fit of C+07.  In fact, broadening of the longer wavelength wing of the silicate profile is one of the changes between the $0.3\leq A_K<1.0$ and $1.0\leq A_K$ extinction curves.  However, H$_2$O absorption around $13\mu$m could also contribute to broadening of that wing.  The difference between the total optical depth (my data) and the relative silicate optical depth (C+07 data) represents the continuum extinction underlying the silicate feature, which is presumably part of the shallow shape of the extinction curve beyond $3\mu$m.  To test whether water ice is associated with this underlying extinction, I plotted the difference between the total and excess optical depths (mine - C+07) against the optical depths of all three ices seen in the extinction curves (Fig. \ref{excess}).  Surprisingly, the correlation between the excess continuum extinction and \emph{all} of the ices is very strong, not just H$_2$O ice.  In addition, H$_2$O and CO$_2$ ice have formation threshold extinctions around $A_K=0.5$ ($\approx3-4$ in $A_V$) \citep{whittet83, murakawa00, bergin05}, which is consistent with the lowest extinctions in our sample.  This 0.5 magnitudes of extinction is also the threshold at which $R_V$ changes from the DISM value of 3.1 to a value of $\approx$ 5 over $0.35-2.2\mu$m along lines of sight in Taurus \citep{whittet01} and other nearby star-forming regions \citep{chapman08}.  

Taken together, these results indicate that ices are associated with the transition from a DISM extinction curve to the molecular cloud curve.  However, the result that \emph{all} of the ices correlate with the underlying continuum exctinction indicates that all of the ice species contribute to the process that creates this continuum extinction.  Since only the water ice libration band at $~13\mu$m could widen the silicate profile, but not the other ice species, another process, such as grain growth could be the main contributor to the widened silicate profile.  Scattering from larger grains could also explain both the shallower $5-8\mu$m and $12-14\mu$m regions.  In fact, Figures 16-18 of \citet{chapman08} show how their IRAC extinction curves compare with the best fitting model curve from a paper in preparation by Pontipiddan et al., a model which was constructed for solid grains with ice mantles \citep{chapman08}, and this model does not provide enough extinction over the $5-8\mu$m and $12-14\mu$m regions to match my $A_K>1$ curve or the \citet{chapman08} and \citet{flaherty07} data.  Given that ices contribute to the shape of the extinction curve, but ice mantles do not match the data well, and we see signs of grain growth, it is likely that we need to consider a different structure in the grains.  A possibility is that that after water ice mantles the grains, they become `sticky' in collisions, forming porous coagulations of smaller grains held together by icy coatings on their surfaces.

\section{Conclusions}

These new curves demonstrate that the shape of the extinction curve changes from a shape close to the DISM extinction curve at $A_K\approx0.5$ to a new shape at higher $A_K>1$, a result which is addressed for the first time here and, independently, in \citet{chapman08}.  That our results, derived with different methods from different data, agree so well is a strong statement in favor of their validity.  Additionally, comparison of the optical depths of the silicate and ice features in these extinction curves indicates that while ices play a significant role in the transition from DISM to molecular cloud extinction, grain growth via coagulation with the ice as a `glue' between the particles is likely to contribute more to the extinction than simple ice mantles alone.  Theoretical models are needed to confirm the role played by ices and grain growth in changing the shape of the extinction curve, but the empirical extinction curves presented here seem appropriate for extinction-correcting the flux of objects with $A_K>0.5$ in molecular clouds.  Future \emph{Spitzer} observations of objects behind dark clouds will hopefully refine our understanding of the change in the shape of the silicate profiles from $0.5<A_K<1$ and to what component or environmental condition this change can be attributed.

\acknowledgments
I would like to thank the anonymous referee, for helpful comments that improved the manuscript, as well as Bill Forrest, Elise Furlan, Ted Bergin, Nuria Calvet, and Dan Watson for their substantial help and patience in discussing the implications of these results and editing several drafts.  This work is based on observations made with the {\it \emph{Spitzer} Space Telescope}, which is operated by the Jet Propulsion Laboratory, California Institute of Technology, under NASA contract 1407. Support for this work was provided by NASA through contract number 1257184 issued by JPL/Caltech, JPL contract 960803 to Cornell University, and Cornell subcontracts 31419-5714 to the University of Rochester.  This publication made use of NASA's ADS Abstract Service as well as the SIMBAD database and Vizier catalog service, operated by the Centre de Données astronomiques de Strasbourg.

\begin{deluxetable}{lcccccccc}
\tabletypesize{\scriptsize}   
\tablewidth{\linewidth}
\tablecaption{Spectral type information for the entire sample \label{SpT_all}}
\tablehead{
\colhead{Name} & \colhead{2MASS id} & \colhead{SpT (III)} & \colhead{PID} & \colhead{$A_K$ $^c$} &\colhead{$\tau_{9.7}$} & \colhead{$\tau_{H_2O(6.0)}$} &  \colhead{$\tau_{`methanol'(6.8)}$} & \colhead{$\tau_{CO_2(15.2)}$}
}
\startdata
\textbf{IC 5146} \\
\hline
Quidust 21-1 & 21472204+4734410 & G0--M4 & 3320 & 2.56$\pm$0.25 & 1.5$\pm$0.13 & 1.2$\pm$0.13 & 1.1$\pm$0.12 & 1.6$\pm$0.23\\
Quidust 21-2 & 21463943+4733014 & G0--M4 & 3320 & 1.23$\pm$0.25 & 0.82$\pm$0.13 & 0.49$\pm$0.13 & 0.52$\pm$0.12 & 0.46$\pm$0.23\\
Quidust 21-3 & 21475842+4737164 & G0--M4 & 3320 & 1.11$\pm$0.25 & 0.73$\pm$0.14 & 0.43$\pm$0.14 & 0.42$\pm$0.12 & 0.38$\pm$0.23\\
Quidust 21-4 & 21450774+4731151 & G0--M4 & 3320 & 0.84$\pm$0.24 & 0.41$\pm$0.13 & 0.29$\pm$0.13 & 0.24$\pm$0.11 & 0.21$\pm$0.23\\
Quidust 21-5 & 21444787+4732574 & G0--M4 & 3320 & 0.73$\pm$0.24 & 0.43$\pm$0.13 & 0.21$\pm$0.13 & 0.14$\pm$0.11 & ...\\
Quidust 21-6 & 21461164+4734542 & G0--M4 & 3320 & 2.2$\pm$0.25 & 1.43$\pm$0.14 & 1.08$\pm$0.14 & 0.92$\pm$0.12 & 1.35$\pm$0.24\\
Quidust 22-1 & 21443293+4734569 & G0--M4 & 3320 & 1.88$\pm$0.25 & 1.27$\pm$0.14 & 0.86$\pm$0.14 & 0.85$\pm$0.12 & 1.39$\pm$0.24\\
Quidust 22-3 & 21473989+4735485 & G0--M4 & 3320 & 0.28$\pm$0.25 & 0.12$\pm$0.14 & 0.04$\pm$0.14 & 0.08$\pm$0.12 & -0.38$\pm$0.24\\
Quidust 23-1 & 21473509+4737164 & G0--M4 & 3320 & 0.73$\pm$0.25 & 0.37$\pm$0.14 & 0.34$\pm$0.14 & 0.26$\pm$0.12 & 0.35$\pm$0.24\\
Quidust 23-2 & 21472220+4738045 & G0--M4 & 3320 & 0.94$\pm$0.25 & 0.62$\pm$0.14 & 0.31$\pm$0.14 & 0.29$\pm$0.12 & 0.32$\pm$0.24\\
\hline
\textbf{Taurus} \\
\hline
Elias 3 & 04232455+2500084 & K2 & 172 & 1.12$\pm$0.08 & 0.7$\pm$0.04 & 0.51$\pm$0.04 & 0.48$\pm$0.04 &  ... \\			
Elias 13 & 04332592+2615334 & K2 & 172 & 1.48$\pm$0.09 & 0.9$\pm$0.04 & 0.62$\pm$0.04 & 0.62$\pm$0.04 &  ... \\			
Elias15$^a$ & 04392692+2552592 & M2 & 172 & 1.85$\pm$0.07 & 1.1$\pm$0.04 & 0.71$\pm$0.04 & 0.62$\pm$0.03 & 0.76$\pm$0.05\\	
Elias 16 & 04393886+2611266 & K1 & 172 & 2.99$\pm$0.06 & 1.7$\pm$0.04 & 1.29$\pm$0.04 & 1.22$\pm$0.03 & 1.55$\pm$0.07\\	
	  &		       &	  & 27 \\
TNS 2$^a$ & 04372821+2610289 & M0 & 172 & 0.81$\pm$0.06 & 0.46$\pm$0.03 & 0.26$\pm$0.03 & 0.19$\pm$0.03 & 0.21$\pm$0.04\\	
TNS 8$^a$ & 04405745+2554134 & K5 & 172 & 2.71$\pm$0.08 & 1.74$\pm$0.04 & 1.14$\pm$0.04 & 1.08$\pm$0.04 & 1.35$\pm$0.05\\	
	  &		      &     & 179 \\
\hline
\textbf{Barnard 59} \\
\hline
B59-bg7 & 17111538-2727144 & G0--M4 & 20604 & 3.58$\pm$0.25 & 2.02$\pm$0.17 & 1.56$\pm$0.17 & 1.45$\pm$0.15 &  ... \\			
B59-bg1 & 17112005-2727131 & G0--M4 & 20604 & >4.94		 & ... & ... & ... & ...\\			
\hline
\textbf{Barnard 68} \\
\hline
Quidust 18-1 & 17224500-2348532 & G0--M4 & 3320 & 0.46$\pm$0.25 & 0.53$\pm$0.14 & 0.09$\pm$0.14 & 0.16$\pm$0.12 & 0.52$\pm$0.24\\
Quidust 19-1 & 17224483-2349049 & G0--M4 & 3320 & 0.61$\pm$0.25 & 0.2$\pm$0.19 & 0.33$\pm$0.19 & 0.13$\pm$0.17 & ...\\
Quidust 20-1 & 17224407-2349167 & G0--M4 & 3320 & 0.76$\pm$0.25 & 0.71$\pm$0.25 & 0.38$\pm$0.25 & 0.32$\pm$0.22 & ...\\
Velucores 1-1 & 17223790-2348514 & G0--M4 & 3290 & 1.39$\pm$0.25 & 0.9$\pm$0.14 & 0.72$\pm$0.14 & 0.58$\pm$0.12 & 0.47$\pm$0.24\\
Velucores 1-2  & 17224511-2348394 & G0--M4 & 3290 & 0.35$\pm$0.24 & 0.21$\pm$0.13 & 0.02$\pm$0.13 & 0.27$\pm$0.12 & -0.31$\pm$0.23\\
Velucores 1-3 & 17224027-2348555 & G0--M4 & 3290 & 1.56$\pm$0.25 & 1.07$\pm$0.15 & 1.07$\pm$0.15 & 0.64$\pm$0.13 & ...\\
Velucores 1-4 & 17224159-2350261 & G0--M4 & 3290 & 2.11$\pm$0.25 & 1.26$\pm$0.35 & 0.74$\pm$0.35 & 2.28$\pm$0.17 & ...\\
\hline
\textbf{Chameleon I} \\
\hline
Quidust 2-1 & 11024279-7802259 & G0--M4 & 3320 & 0.73$\pm$0.25 & 0.45$\pm$0.14 & 0.15$\pm$0.14 & 0.17$\pm$0.12 & 0.19$\pm$0.23\\
Quidust 2-2 & 11055453-7735122 & G0--M4 & 3320 & 1.63$\pm$0.25 & 1.09$\pm$0.14 & 0.66$\pm$0.14 & 0.6$\pm$0.12 & 0.73$\pm$0.24\\
Quidust 3-1 & 11054176-7748023 & G0--M4 & 3320 & 1.03$\pm$0.25 & 0.65$\pm$0.14 & 0.44$\pm$0.14 & 0.38$\pm$0.12 & 0.48$\pm$0.24\\
\hline
\textbf{Serpens} \\
\hline
CK2 & 18300061+0115201 & K4$^b$ & 40525 & 6.27$\pm$0.04 & 4.27$\pm$0.1 & 3.04$\pm$0.1 & 3.01$\pm$0.09 & 3.3$\pm$0.17\\
SVS76 Ser 9 & 18294508+0118469 & G0--M4 & 172 & 2.23$\pm$0.25 & 1.54$\pm$0.15 & 0.88$\pm$0.15 & 0.84$\pm$0.13 & 0.95$\pm$0.18\\
	  &		      &     & 179 \\
SSTc2d182852.7 & 18285266+0028242 & G0--M4 & 179 & 4.94$\pm$0.34 & 3.62$\pm$0.18 & 2.5$\pm$0.18 & 2.43$\pm$0.15 & 2.92$\pm$0.22\\
+02824 \\
\hline
\enddata

\tablecomments{
$^a$ object from \citet{shenoy08}, $^b$ SpT from \citet{knez05}, $^c$ The conversion factor from $A_K$ to $A_V$, assuming $R_V=5.0$, is $A_V/A_K=7.75$   \\
}

\end{deluxetable}

\begin{deluxetable}{ccc}
\tabletypesize{\scriptsize}   
\tablewidth{\linewidth}
\tablecaption{Final, composite extinction curves constructed from the extinction curves derived here and others from the literature as described in the $2^{nd}$ paragraph of Section 3.1 for $0.3\leq A_K<1$ and $1\leq A_K<7$. The wavelengths over the mid-infrared are sampled to the SL and LL \emph{Spitzer} IRS modules. \label{exttab}}
\tablehead{
\colhead{wavelength} & \colhead{$A_{\lambda}/A_K$} & \colhead{$A_{\lambda}/A_K$}}
\startdata
($\mu$m) & $0.3\leq A_K<1$ & $1\leq A_K<7$\\
\hline
5.19 & 4.13E-1 & 4.89E-1\\
5.22 & 4.10E-1 & 4.87E-1\\
5.25 & 4.08E-1 & 4.85E-1\\
5.28 & 4.06E-1 & 4.83E-1\\
5.31 & 4.04E-1 & 4.81E-1\\
5.34 & 4.02E-1 & 4.79E-1\\
5.37 & 4.00E-1 & 4.77E-1\\
5.40 & 3.98E-1 & 4.75E-1\\
5.43 & 3.96E-1 & 4.73E-1\\
5.46 & 3.94E-1 & 4.71E-1\\
5.49 & 3.92E-1 & 4.70E-1\\
5.52 & 3.90E-1 & 4.68E-1\\
\hline
\enddata

\tablecomments{Full table is available in machine-readable format in the electronic version of this article.}

\end{deluxetable}

\end{document}